\documentclass[conference]{IEEEtran}
\usepackage{cite}
\usepackage{amsmath,amssymb,amsfonts}
\usepackage{algorithmic}
\usepackage{algorithm2e}
\usepackage{graphicx}
\usepackage{textcomp}
\usepackage{xcolor}
\usepackage{multirow}
\usepackage{hhline}
\usepackage{mathtools}
\usepackage{amsmath}
\usepackage{hyperref}
\usepackage{color}
\usepackage{epsfig}
\usepackage{epstopdf}
\usepackage{dblfloatfix} 
\usepackage{graphicx}
\def\BibTeX{{\rm B\kern-.05em{\sc i\kern-.025em b}\kern-.08em
    T\kern-.1667em\lower.7ex\hbox{E}\kern-.125emX}}
\begin{document}

\title{Approximate Logic Synthesis Using BLASYS \\
{\normalsize (WOSET-2019)}}

\author{\IEEEauthorblockN{Jingxiao Ma, Soheil Hashemi, and S. Reda}
\IEEEauthorblockA{\textit{School of Engineering, Brown University, Providence RI 02912} \\
}
}

\maketitle

\begin{abstract}
Approximate computing is an emerging  paradigm where design accuracy can be traded  for improvements in design metrics such as design area and power consumption. In this work, we overview our open-source tool, BLASYS, for synthesis of approximate circuits using Boolean Matrix Factorization (BMF). In our methodology the truth table of a given circuit  is approximated using BMF to a controllable approximation degree, and the results of the factorization are used to synthesize the approximate circuit output. BLASYS scales up the computations to large circuits through the use of  partition techniques, where an input circuit is partitioned into a number of interconnected subcircuits and then a design-space exploration technique  identifies the best order for subcircuit approximations. BLASYS leads to a graceful trade-off between accuracy and full circuit complexity as measured by design area. Using an open-source design flow, we extensively evaluate our methodology on a number of benchmarks, where we demonstrate that the proposed methodology can achieve on average 48.14\% in area savings, while introducing an average relative error of 5\%. 

\end{abstract}


\maketitle

\section{Introduction}
\label{sec:introduction}

Approximate computing techniques trade off accuracy with improvements in design area and power consumption \cite{reda19}. Approximate computing is attractive in applications domains that are inherently tolerant to errors such as machine learning, computer vision, computer graphics, and signal processing. A central issue in approximate computing is how to automatically synthesize an approximate circuit from an input circuit \cite{LiL15,salsa,aslan14,sasimi,Miao13,Nepal14,Lee17,Nepal19}.

This paper overviews the public release of BLASYS, which is a tool to synthesize approximate circuits \cite{hashemi2018blasys,hashemi19}. BLASYS is based on Boolean Matrix Factorization (BMF), where the truth table of a circuit is represented as a matrix that is factorized into two smaller Boolean matrices that are then synthesized into the approximate circuit. BMF is a derivative of non-negative matrix factorization (NNMF) \cite{Lee99,Miettinen14}. The non-negativity constraints on the factorization arise in physical domains, such as computer vision and document clustering~\cite{XuL03}. Recent advances in the mathematical community extend NNMF techniques to Boolean matrices, where matrix operations are carried in $GF(2)$~\cite{Miettinen14}. The use of BMF creates a solid foundation for approximate logic synthesis, and enables systematic trade-off between design complexity and Quality of Results (QoR).  Since enumerating the truth table is limited to only circuits of moderate number of inputs, BLASYS partitions a large-input circuit into smaller subcircuits, that are then  approximated individually using BMF. We devise a greedy design space exploration method that identifies a good ordering for the subcircuits to approximate together with the approximation degree for each subcircuit. Our tool uses a full stack of open-source tools, including LSOracle for circuit partitioning \cite{lsoracle}, Yosys for Verilog parsing \cite{wolf2016yosys}, iVerilog for simulation and QoR estimation \cite{williams2006icarus}, and ABC for logic synthesis \cite{brayton2010abc}. We evaluate our approach on a large a number of circuits from the EPFL benchmark set \cite{Amarù:207551}. We show that our approach is able to trade-off accuracy with circuit area  in a graceful manner. 

The organization of this paper is as follows.  We discuss the details of BLASYS  in Section \ref{sec:methodology}, where we describe the basic idea of using BMF algorithms to approximate logic circuits, and then show how to scale our proposed method to larger circuits. We describe BLASYS tool-chain flow in Section \ref{sec:flow}. We provide results from the public release of BLASYS in  Section~\ref{sec:results}. Finally, we summarize our main conclusions and directions for future work in Section~\ref{sec:conclusions}.

\section{Proposed Methodology}
\label{sec:methodology}
\begin{figure}[b!]
\vspace{-0.1in}
	\begin{center}
		\includegraphics[scale=0.25]{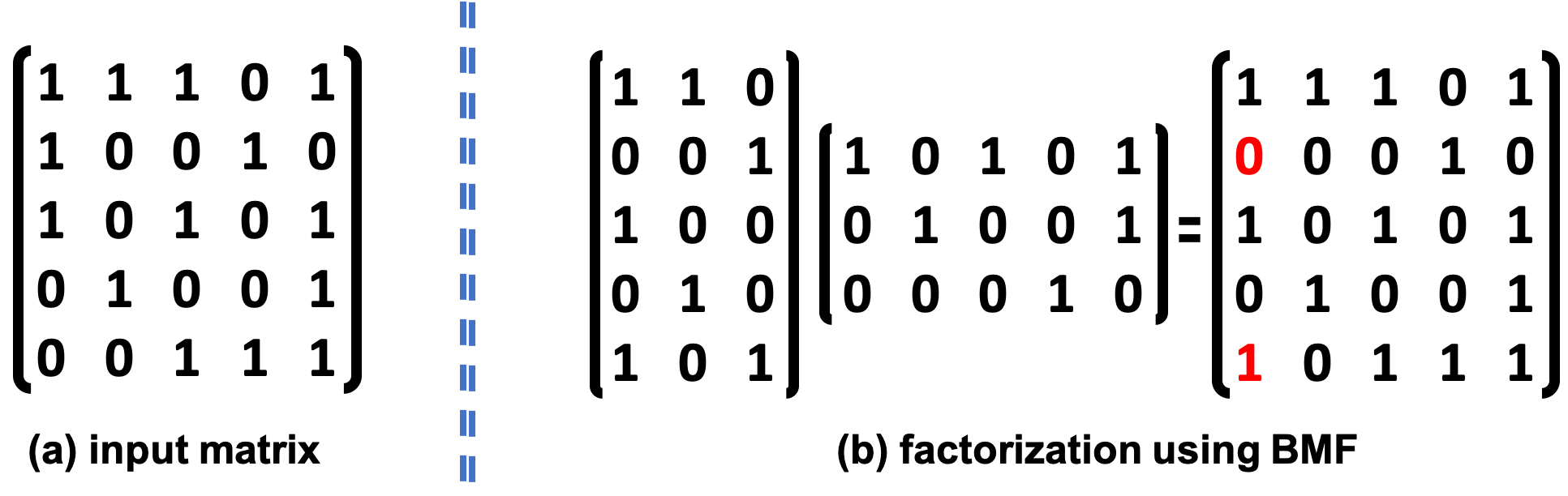}
		\vspace{-0.1in}
		\caption{BMF example.}
		\label{fig:nmf}
	\end{center}
    \vspace{-0.1in}
\end{figure}

Non-negative matrix factorization (NNMF) is a factorization technique where a $k\times m$ matrix $\mathbf{M}$ is factored into two non-negative matrices: a $k\times f$ matrix $\mathbf{B}$, and an $f \times m$ matrix $\mathbf{C}$, such that $\mathbf{M} \approx \mathbf{B} \mathbf{C}$ \cite{Lee99}.   NNMF has been extended to Boolean matrices where elements of all matrices are restricted to Boolean values. In this case, multiplications can be performed using logical AND, and additions are performed using logical OR  \cite{Miettinen14}.
Figure \ref{fig:nmf} provides an example of BMF over $GF(2)$.

\begin{figure}[t!]
	\begin{center}
		\includegraphics[scale=0.27]{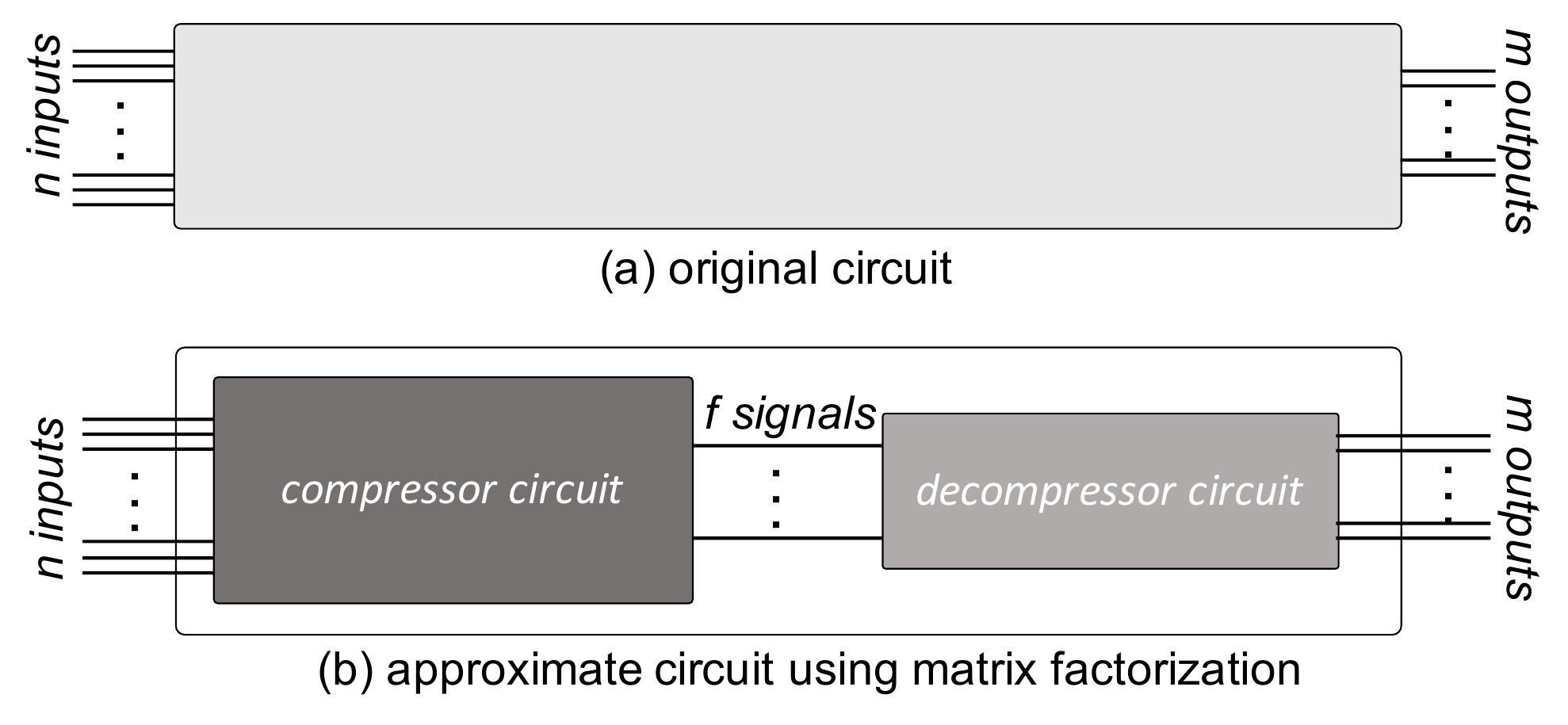}
		\caption{Generating approximate circuits using BMF.}
		\label{fig:bmf}
	\end{center}
    \vspace{-0.1in}
\end{figure}

In our proposed approach, a multi-output logic circuit with $k$ inputs and $m$ outputs is first evaluated to generate its truth table. The truth table, represented by $\mathbf{M}$, is then given as input to a BMF algorithm together with the target factorization degree $1\leq f < m$, to produce the two factor matrices $\mathbf{B}$ and $\mathbf{C}$. Matrix $\mathbf{B}$ is then given as the input truth table to a logic synthesis tool to generate a $k$ input, $f$ output circuit, which we refer to as the {\it compressor circuit}. Note that the compressor matrix is simply the truth table of a circuit with the same number of inputs as the original circuit but with fewer ($f$ to be exact) output signals. Therefore, it can easily be mapped to logic. These $f$ outputs from the compressor circuit are then combined by the {\it decompressor circuit} according to the $\mathbf{C}$ matrix using a network of OR gates  to generate the approximate $m$ outputs as illustrated in Figure~\ref{fig:bmf}. The compressor and decompressor circuits can then be optimized by logic tools as one circuit to remove any logic redundancies. Using this methodology, any arbitrary circuit can be approximated by forcing the circuit to compress as much information as possible in $f$ intermediate signals and then decompress them using simple OR  gates.

Calculating the BMF is limited by computational complexity as one needs to generate the truth table for every possible input and state combination. Furthermore, BMF is a NP-hard problem, and most algorithms in the literature are heuristics \cite{Lee99,Miettinen14}. To address scalability, we partition a large circuit into a number of subcircuits, each with a maximum number of $k$ inputs. $k$ is chosen to make the runtime of the factorization algorithm tractable and to meet memory requirements. In our case we use a modified approach of the hypergraph partitioning tool KaHyPar \cite{Amarù:207551} as part of LSOracle \cite{}. We  then proceed to approximate each subcircuit individually. To evaluate the QoR of an approximate subcircuit, we first substitute a partitioned  subcircuit by its approximate counterpart, and then evaluate the QoR of the entire modified circuit using the supplied testbenches. In our testbenches,  we use Monte Carlo sampling to estimate the QoR of the approximate version of the entire circuit. 

Partitioning a large circuit can yield tens of interconnected subcircuits. The order of processing the subcircuits and  the target factorization degree for each subcircuit impact the results. BLASYS performs iterative design space exploration to identify a good ordering.  At each iteration, BLASYS evaluates the impact of approximating each subcircuit by reducing its factorization degree by one, and assessing the resulting reduction in total circuit area  and  loss in QoR. The approximate subcircuit that leads to the smallest ratio of error to area reduction is then chosen and incorporated into the full circuit. The iterative process is then repeated as long as the error is above the set threshold or all subcircuits are approximated to the minimum possible factorization degree, i.e.,  $f=1$. 


\section{BLASYS Tool Chain}
\label{sec:flow}

\begin{figure}[t]
\centering
\scalebox{0.300}{
\epsfig{file=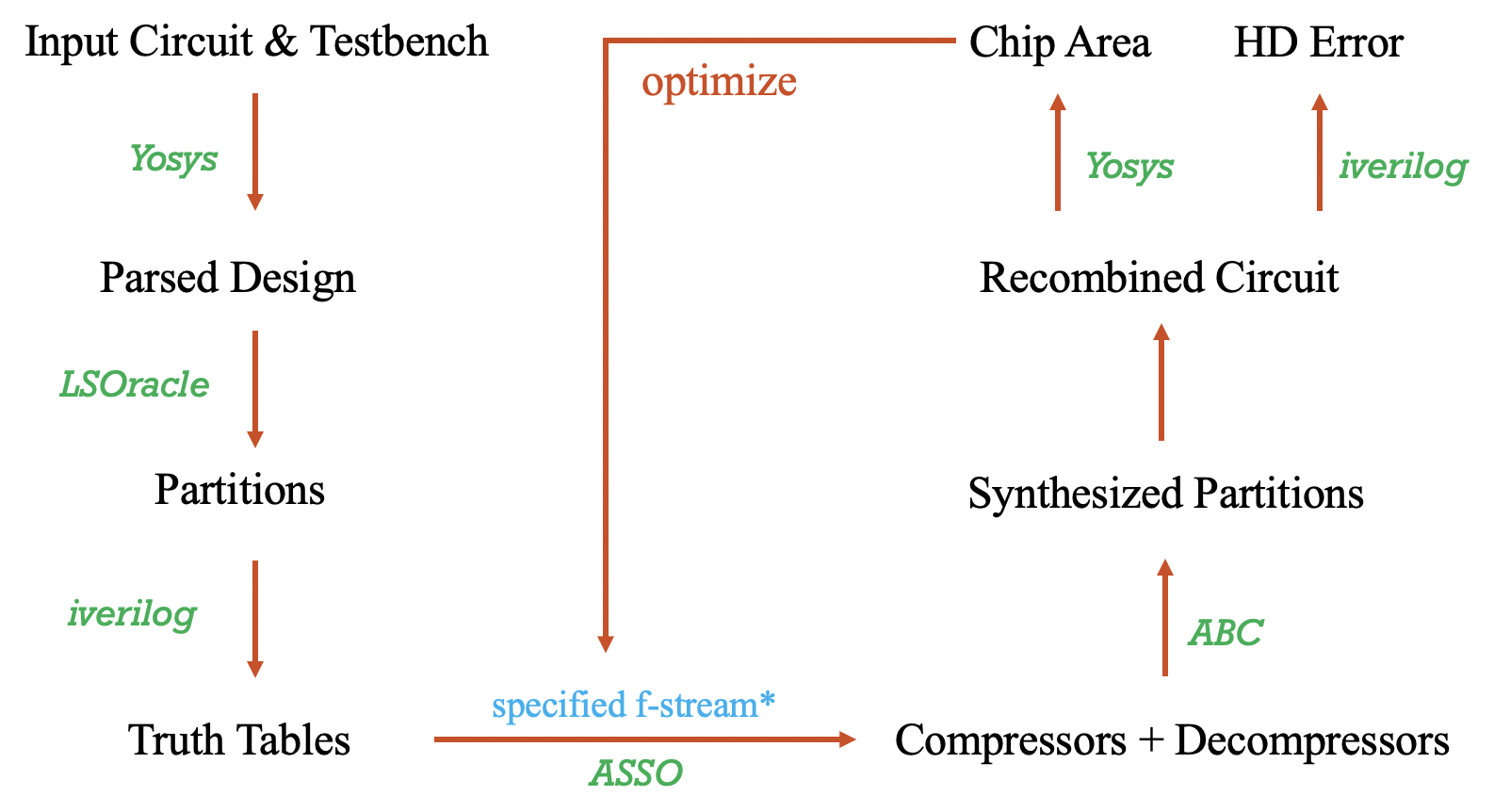}}
\vspace{-0.10in}
\caption{BLASYS flow.}
\vspace{-0.20in}
\label{fig:blasys}
\end{figure}


Figure \ref{fig:blasys} prvoides the flow of BLASYS. To begin with, an input circuit is parsed with Yosys \cite{wolf2016yosys} to estimate initial chip area, where ASAP 7nm design library is used. Using  the provided testbench, BLASYS simulates the input circuit and store original results for QoR estimation. The simulation tool we used is Icarus Verilog \cite{williams2006icarus}.

Next, LSOracle \cite{lsoracle} is used to partition the input circuit to multiple subcircuits, each of which has a similar size. BLASYS partitions an input circuit until all partitions have less than 16 inputs. BLASYS then creates a testbench that generates the truth table for each subcircuit.  We use the ASSO algorithm \cite{Miettinen14} to perform BMF on each truth table based on a vector called {\it f-stream}, which consists of factorization degree for each subcircuit, which is determined by the design space exploration method as discussed in Section \ref{sec:methodology}. As a result, each truth table is factorized into a compressor and decompressor. BLASYS calls ABC \cite{brayton2010abc} to synthesize the compressor matrix  to a circuit and uses a network of logic OR to represent decompressor. Thus, an approximated version of the input circuit can be obtained by recombining all approximated subcircuits. Afterwards, BLASYS calls Yosys to estimate the chip area of the approximate circuit and executes a simulation using the input testbench. From the  original and approximated simulation results, QoR can be estimated by measuring the Hamming Distance Error, which is the number of output bit flips in the results of the testbench simulation divided by the total number of output bits. The area reduction ratio and QoR are used to optimize f-stream iteratively and greedily as mentioned in Section \ref{sec:methodology}. To bring the the runtime of BLASYS to practical levels for  large circuits, we  implement a parallel mode, which allows our flow to work on multiple designs simultaneously using multiple cores.

\section{Experimental Results}
\label{sec:results}

\begin{figure}[t!]
\centering
\scalebox{0.55}{
\epsfig{file=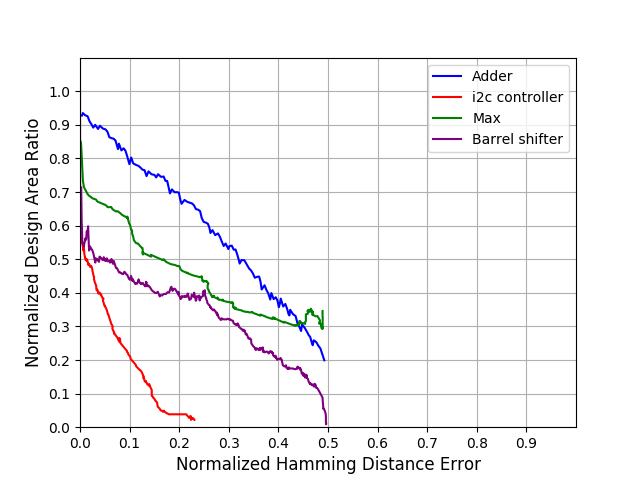}}
\vspace{-0.10in}
\caption{Trade-off between design area and QoR for 4 benchmarks.}
\vspace{-0.20in}
\label{fig:lw_vs_nw}
\end{figure}

In our experiments we consider a number of benchmarks from the EPFL benchmark set \cite{Amarù:207551}, and set $k\leq 16$. These numbers are simply chosen as they provide a balanced trade-off between truth table complexity and number of subcircuits. To generate the testbenches, we use Monte Carlo simulation with 10,000 randomly generated input test vectors. For QoR, we measure the normalized Hamming Distance Error by ratio of flipped bits. Figure~\ref{fig:lw_vs_nw} shows trade-off between accuracy and design area on four benchmarks based on our design-space exploration technique. Although global optimum is not guaranteed, our results show significant area reduction within small HD error interval. This is beneficial since relatively small amount of error makes more sense in practice. We tabulate the results from all our evaluated benchmarks in  Table~\ref{table:apps}, where we report area reduction ratio at error threshold 5\% and 10\% in Table~\ref{table:apps}. The second column in the table provides I/O information about benchmarks, and the third column provide the number of partitioned subcircuits. Based on the circuit, benefits of 11\%-71\% can be achieved within 5\% HD error, which leads to 48.14\% chip area reduction on average. The number increases to 60.64\% with 10\% HD error metric. As the figure and table demonstrate, BLASYS achieves a large amount of area reduction within small HD error.

The runtime of our tool is dominated by simulation process. For example, it takes around 9.4 seconds to simulate Max circuit on 10,000 test cases with iVerilog. Due to the greedy optimization scheme, total times of simulation grows quadratically with number of partitions. Thus, choosing a proper partitioning is vital. Our modified ASSO algorithm works efficiently on subcircuits with $k\leq 16$. With 16 inputs, ASSO takes around 0.6 second to perform BMF.

\begin{table}[t!]
  \footnotesize
  \centering
    \caption{Improvement in design area as a function of allowed maximum Hamming Distance (HD) error.}
    \vspace{-0.1in}
  \begin{tabular}{l|c||c|c|c}
& & Number of& \multicolumn{2}{c}{Area Imp. (\%)}\\
\cline{4-5}
   benchmark & I/O  & Partitions &  5\% HD   & 10\% HD   \\

     \hline \hline
 Adder & 256/129 & 20 & 11.09 & 21.71  \\ \hline
  Alu control unit & 7/26 & 5 & 51.71 & 71.03 \\ \hline
 Barrel shifter & 135/128& 25 & 49.41 & 55.95 \\ \hline
 Coding-cavlc & 10/11& 50 & 71.67 & 90.56 \\ \hline
 i2c controller & 147/142 & 100 & 63.86 & 78.66 \\ \hline
 int2float converter & 11/7 & 5 & 35.32 & 47.27 \\ \hline
 Max & 512/130 &  150 & 53.90 & 59.33 \\ \hline \hline
 \multicolumn{3}{c||}{\textbf{Average}} & 48.14 & 60.64\\
  \hline

  \end{tabular}
  \vspace{-0.2in}
  \label{table:apps}
\end{table}

\section{Conclusions}
\label{sec:conclusions}

In this paper we overviewed the public release of BLASYS. We have demonstrated that BLASYS can synthesize approximate version for arbitrary large circuits in Verilog. BLASYS partitions the circuits using hypergraph partitioning techniques and then proceeds to approximate the individual circuit partition, while taking into account the impact on global area and QoR. BLASYS leads to a systematic approach to trade-off accuracy with circuit complexity. Our tool is available at \url{http://github.com/scale-lab/blasys}.\\

\vspace{-0.1in}
\noindent\textbf{Acknowledgments:} {\small This work is partially supported by NSF grant 1814920 and DoD ARO grant W911NF-19-1-0484.}

\vspace{-0.10in}
\bibliographystyle{plain}
\bibliography{sreda,refbib2,jma} 

\end{document}